\documentclass[12pt,preprint]{aastex}

\slugcomment{Not to appear in Nonlearned J., 45.}

\shorttitle{H$_2$O Maser Gas and a Dense Plasma in NGC 1052}
\shortauthors{Sawada-Satoh et al.}

\begin{document}


\title{Positional Coincidence of H$_2$O Maser
and a Plasma Obscuring Torus in Radio Galaxy NGC 1052}


\author{Satoko Sawada-Satoh \altaffilmark{1,2},  
Seiji Kameno\altaffilmark{3},  Kayoko Nakamura \altaffilmark{3}, 
 Daichi Namikawa \altaffilmark{3},  
 Katsunori M. Shibata \altaffilmark{4},  
 Makoto Inoue \altaffilmark{4}}

\altaffiltext{1}{Department of Physics, Faculty of Science, Yamaguchi University, 
1677-1 Yoshida, Yamaguchi, 753-8512 Japan}
\email{sss@yamaguchi-u.ac.jp}
\altaffiltext{2}{Academia Sinica Institute of Astronomy and Astrophysics, 
P.O. Box 23-141, Taipei 10617, Taiwan}
\altaffiltext{3}{Department of Physics, Faculty of Science, Kagoshima University, 
1-21-35 Korimoto, Kagoshima, 890-0065, Japan}
\altaffiltext{4}{National Astronomical Observatory, Mitaka, Tokyo 181-8588, Japan.}


\begin{abstract}
We present multi-frequency simultaneous VLBA observations 
at 15, 22 and 43 GHz 
towards the nucleus of the nearby radio galaxy NGC~1052.  
These three continuum images reveal a double-sided jet structure, 
whose relative intensity ratios imply 
that the jet axis is oriented close to the sky plane. 
The steeply rising spectra at 15--43 GHz at the inner edges of the jets 
strongly suggest that synchrotron emission 
is absorbed by foreground thermal plasma.  
We detected H$_2$O maser emission in the velocity range of 1550--1850 
km~s$^{-1}$, which is redshifted by 50--350 km~s$^{-1}$ 
with respect to the systemic velocity of NGC~1052. 
The redshifted maser gas appears projected against both sides of the jet, 
in the same manner as the H{\scriptsize I} seen in absorption. 
The H$_2$O maser gas are located where the free-free absorption opacity is large. 
This probably imply that the masers in NGC~1052 are associated with 
a circumnuclear torus or disk as in the nucleus of NGC~4258. 
Such circumnuclear structure can be the sence of accreting onto the central engine. 
\end{abstract}





\section{Introduction}

NGC~1052 is a nearby ($z=0.0049$ ; Knapp et~al. 1978) 
radio galaxy whose nuclear activity is classified 
as LINER (e.g. Gabel et al. 2000).
This galaxy hosts a well defined double-sided radio jet 
elongated by several pc with P.A. $\sim65^{\circ}$.  
The jet emanates from the nucleus and can be traced out to kilo-pc-scales 
(Cohen et al. 1971; Wrobel 1984; Jones 1984, Kellermann et al. 1998;
Kameno et al. 2001; Vermeulen et al. 2003; Kadler et al. 2004b).
Radio observations of the low-luminosity active galactic nucleus (AGN) 
of NGC 1052 with Very Long Baseline Interferometry (VLBI) at multiple frequencies have revealed the presence of a dense circumnuclear structure, 
which obscures the very center of this elliptical galaxy 
(Kellermann et al. 1998; Kameno et al. 2001; Vermeulen et al. 2003;
Kadler et al. 2004). 
Proper motion between the two sides of the jet and structural evolution 
have been detected in this galaxy. 
VLBI observations showed that the proper motion between the two sides of the jet  
has an apparent velocity of (0.26$\pm$0.04)$c$  from 1995 to 2001 
(e.g. Kellermann et~al. 1998, Vermeulen et al. 2003). 
A gap between the eastern and western jets, where the nucleus is supposed to be, 
can be seen until 1999 
(Claussen et~al. 1998, Kellermann et~al. 1998, Vermeulen et~al. 2003). 
In 2000, however, a nuclear component appeared 
between the eastern and western jets (Kameno et al. 2001).  
These authors found a central condensation of the plasma which covers about 
0.1~pc and 0.7~pc of the approaching and receding jets, respectively.  
They proposed a pc-scale circumnuclear torus 
model for NGC~1052. 
X-ray spectra also imply a high column density 
of $10^{22}$ cm$^{-2}$ to $10^{23}$ cm$^{-2}$
toward the center, and support the presence of a dense gas torus
(Guainazzi and Antonelli 1999; Weaver et al. 1999; Kadler et al. 2004a). 

The center of NGC~1052 harbors a luminous H$_2$O megamaser, 
which is redshifted by 50--350 km~s$^{-1}$ with respect to 
the systemic velocity of the galaxy 
(1491~km~s$^{-1}$; de Vaucouleurs et al. 1991). 
The spectral profile typically shows a broad velocity 
width of $\sim$100~km~s$^{-1}$ (FWHM) (Braatz et al. 1994, Braatz et al. 1996, Braatz et al. 2003). 
Past VLBI images reveal that 
H$_2$O maser gas with the velocity range of 1585--1685~km~s$^{-1}$ 
is distributed along the western jet, 
0.05--0.1~pc shifted to the west from the gap between the double-side jets in November 1995. 
Excitation by shocks 
into the dense molecular clump which lies in or around the radio jet, 
or amplification of the radio continuum emission of the jet 
by foreground molecular clouds,  were suggested by Claussen et al. (1998). 
On the other hand, 
Kameno et al. (2005) presented the circumnuclear torus model to explain the time variability of 
the H$_2$O maser emission. 
Relevant to the interpretation of the H$_2$O maser emission line, 
several other absorption lines are also found toward the center of NGC~1052 
(H{\scriptsize I} for van Gorkom et al. 1986; OH for Omar et al. 2002; 
HCO$^{+}$, HCN and CO for Liszt and Lucas 2004). 

In order to confirm the positional relation 
between the H$_2$O maser gas and the proposed circumnuclear torus, 
we observed the continuum and maser emissions in the nucleus 
of NGC~1052 with the Very Large Baseline Array (VLBA). 
We adopt $z=0.0049$ (Knapp et~al. 1978) which corresponds 
to a distance $D=$~20~Mpc to NGC~1052 
assuming $H_{0}=75$~km~s$^{-1}$~Mpc$^{-1}$ and $q_0=0.5$.  
Hence 1 mas corresponds to 0.095~pc.

\section{Observations and data reductions}

The observations toward NGC~1052 were carried out at 15, 22 and 43~GHz 
on July 24 2000 with all ten antennas of the VLBA. 
On source times of NGC~1052 at 15, 22 and 43~GHz were, 
170, 190 and 130 minutes, respectively. 
We observed 3C454.3 and 3C84 for phase calibrations and bandpass calibrations. 

The data were recorded in left circular polarization over four 8-MHz 
IF channels, and were divided into 256 spectral channels for each IF channel. 
The continuum emissions were measured with all four IF channels 
at 15 and 43 GHz. 
At 22 GHz, the 
LSR velocities of 1000, 1200, 1650 and 1750 km~s$^{-1}$ were 
each centered on one of the IF channels.   
The two IF channels centered on 1650 and 1750 km~s$^{-1}$,  covered 
the velocity range from 1560 to 1840 km$^{-1}$, 
where the H$_2$O maser emissions have been detected. 
The other two IF channels were used to observe the line-free continuum emission. 
The correlation process was done using the NRAO VLBA correlator. 

Data reduction including calibration, data flagging, fringe fitting 
and imaging utilized using the NRAO AIPS package.  
3C 454.3 and 3C 84 were used as clock-offset and bandpass calibrators. 
All spectral channels at 15 and 43 GHz and the line-free spectral  
channels at 22 GHz were integrated into  a single continuum channel 
for each band. 
Fringe fitting was conducted using the single continuum channel for each band. 
The hybrid imaging process, which involved iterative imaging with CLEAN and 
self calibration, was done at all frequency bands. 
At 22 GHz, 
the solutions of fringe fitting and self calibration obtained 
from the 22-GHz continuum channel were applied to line channels of H$_2$O 
maser emission. 

The overlapping channels of the two IFs with the H$_2$O maser line 
were removed and the two IFs were joined together using the AIPS task UVGLU. 
Channel maps of H$_2$O maser emission were made every 6.74~km~s$^{-1}$, 
averaging every 10 spectral channels. 
The continuum emission of the line channels were subtracted 
from the each spectral channel map in the ${\it u, v}$ plane 
using the 22-GHz continuum map in the AIPS task UVSUB. 
The relative positional accuracy of a maser spot ranged over 0.02--0.18 mas, 
depending on the signal-to-noise ratio and the spatial structure of the spot. 
The synthesized beam size and rms level 
for the 15, 22 and 43 GHz images are 
given in table~1.





\section{Results}

\subsection{Jet knot identifications and alignment} 

\begin{figure}
\epsscale{.90}
\plotone{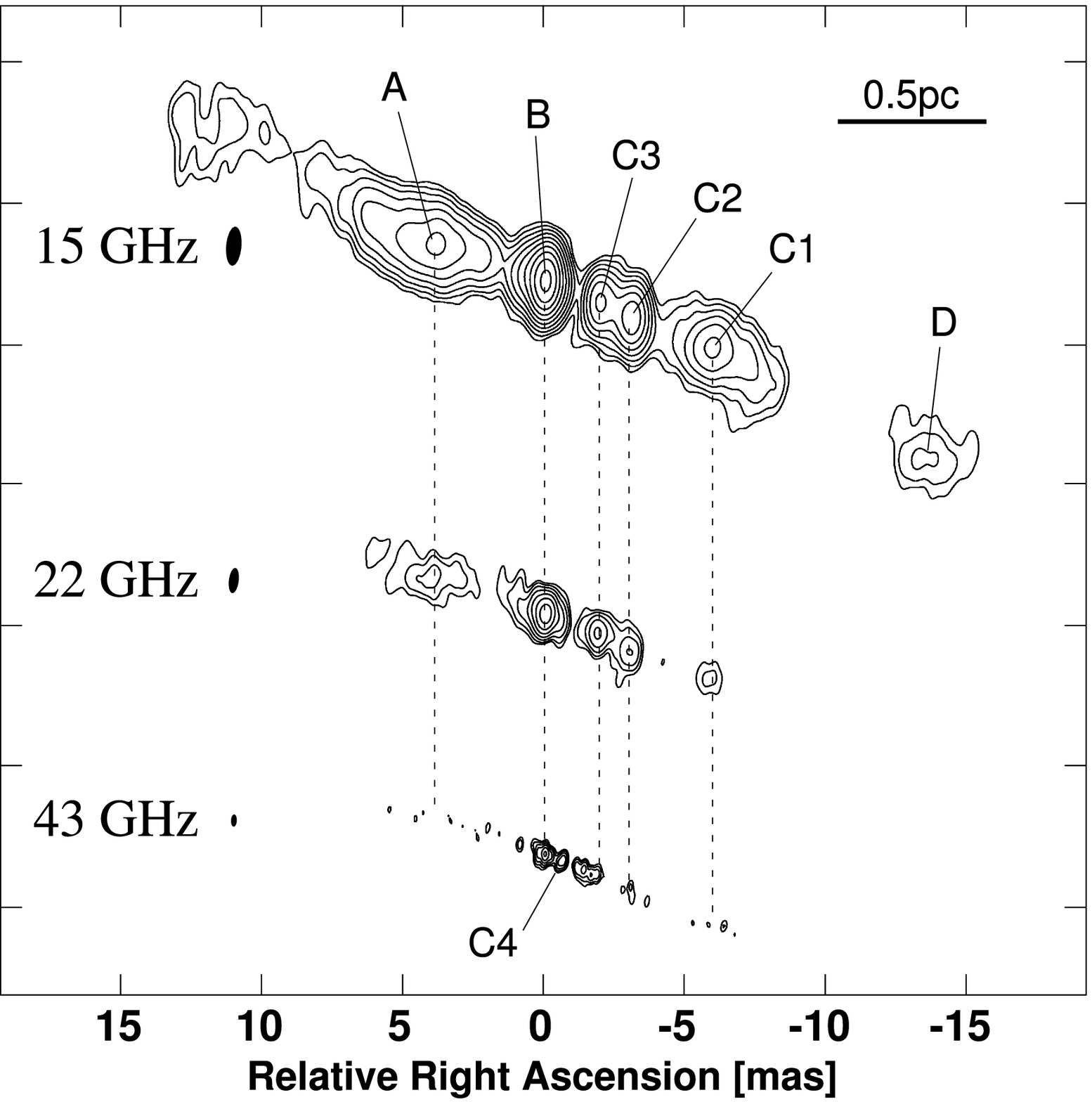}
\caption{The aligned images of NGC~1052 at 15, 22 and 43 GHz 
with VLBA.  
The synthesized beam sizes (FWHM) are 1.3$\times$0.49 mas in P.A.$= -4.8^{\circ}$, 0.86$\times$0.32 mas in P.A.$=-7.1^{\circ}$, 
0.39$\times$0.17 mas in P.A.$=-2.8^{\circ}$, respectively, 
as shown in the left of each image.
The contours start at $3\sigma$ level, increasing by a factor of 2, 
where $\sigma=$ 0.24, 1.07 and 1.45~mJy~beam$^{-1}$, respectively 
at 15, 22 and 43~GHz. }
\label{contmap}
\end{figure}

Figure~\ref{contmap} displays the aligned continuum images toward the nucleus 
of NGC~1052 at 15, 22 and 43~GHz. 
It shows the double-sided jet structure which consists of several components. 
Here we give labels to the eastern jet knot (A), 
the brightest knot (B), and the western jet knots 
(C1, C2, C3 and D), respectively, 
following Kameno et al. (2001) and Kadler et al. (2004b).  
The extended structures of knots A, C1 and D become fainter at higher frequency bands.
Knots B, C2 and C3 are resolved into several components at 22 and 43 GHz. 
The 43~GHz image reveals another knot (C4) located between B and C3. 
In the 43~GHz image, knot A has poorly defined morphology, 
and knot C1 is split into several faint peaks. 
The low resolution image of 43~GHz obtained by restoring 
with a 1.30$\times$0.49 mas beam 
shows the knots C1 and C2 much more clearly.  
The Gaussian fitted parameters of the knots for each frequency are listed in 
table~2. 

The three images at different frequency include uncertainty 
in absolute positional information through the self-calibration process. 
For alignment of these images, we used the relative positions 
($x^{\nu}_k$, $y^{\nu}_k$) of B, C1 and C2, 
which are clearly seen in the restored images with a 1.30$\times$0.49 mas beam 
at 15, 22 and 43 GHz, where $k$ and $\nu$ are the knot id (B, C1, C2) 
and a frequency, respectively. 
Then we derived relative offsets ($\delta x^{\nu}_k$, $\delta y^{\nu}_k$)
to minimize the positional residuals $\chi^{2}_{\nu}$ defined as 
\begin{equation}
\chi^{2}_{\nu}= \sum_{k} \biggl[ 
\frac{(x^{\nu}_k - \delta x^{\nu} - x^{\nu_0}_{k})^2}{\sigma^2_{x^{\nu}_k}+ \sigma^2_{x^{\nu_0}_k}}
+
\frac{(y^{\nu}_k - \delta y^{\nu} - y^{\nu_0}_{k})^2}{\sigma^2_{y^{\nu}_k}+ \sigma^2_{y^{\nu_0}_k}}
\biggr]
\end{equation}
where $\sigma^2_{x^{\nu}_k}$ and $\sigma^2_{y^{\nu}_k}$ are standard positional errors of knot $k$ 
at frequency $\nu$ from the Gaussian fitting, 
and $\nu_0$ is the frequency of the reference image (e.g. Kameno et al. 2001). 
We choose 22~GHz as the reference frequency $\nu_0$. 
Finally, we could overlay the image at 15~GHz and 43~GHz with the image at 22~GHz,  
with positional errors of ($\pm$0.05, $\pm$0.04) mas and ($\pm$0.05, $\pm$0.09) mas, in R.A. and DEC respectively.

\subsection{Free-Free Absorption Opacity Distribution}

\begin{figure}
\epsscale{.90}
\plotone{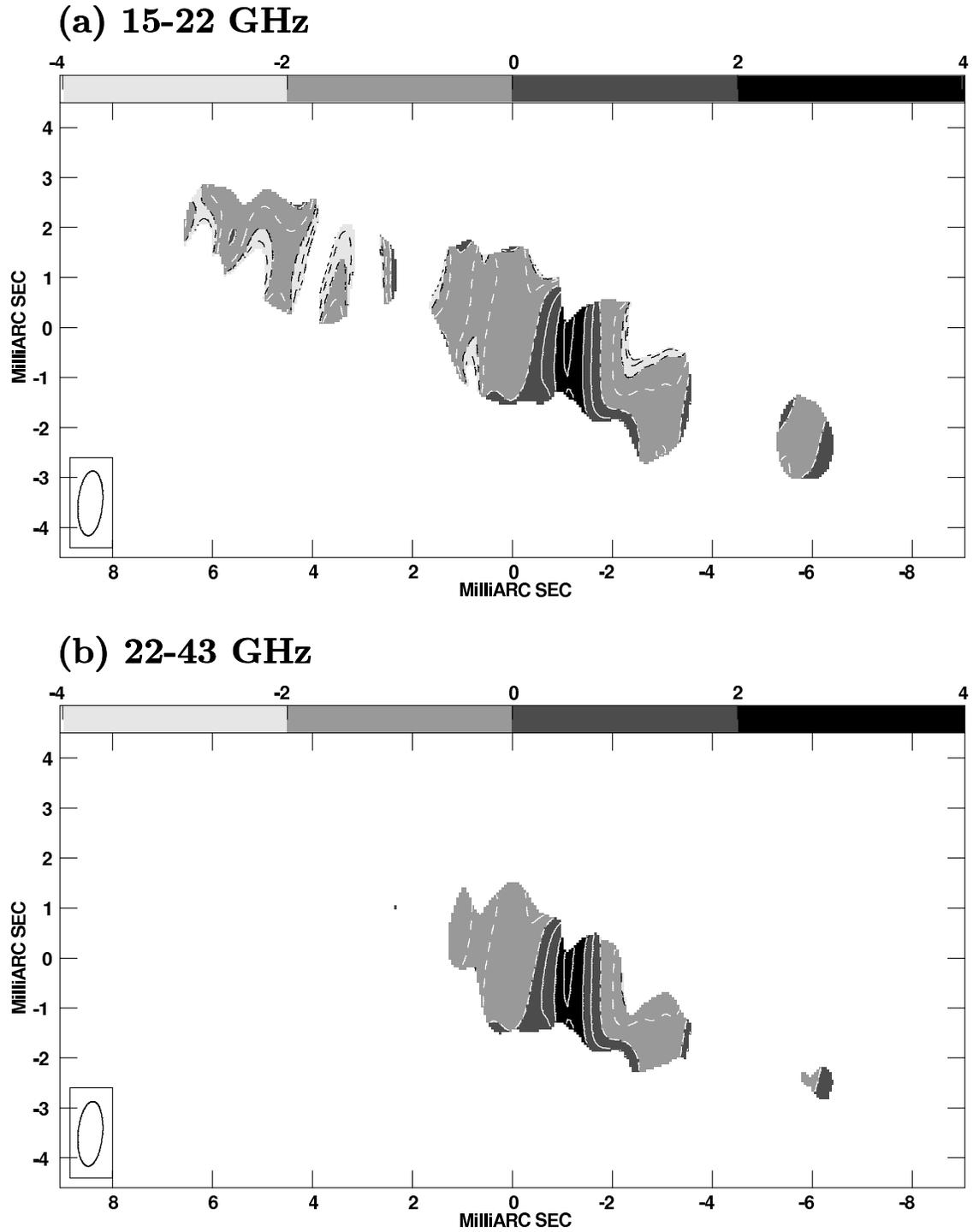}
\caption{Spectral index images of (a) 15--22~GHz, and (b) 22--43~GHz.
The all images were restored with a same beam size 
of 1.30$\times$0.49 mas, as shown in the lower left corner.  
The interval of contour levels is 1, and the dashed lines indicate negative.   }
\label{spcindex}
\end{figure}

After restoring with the same beam size (1.30$\times$0.49 mas), 
we obtained spectral index images using the restoring images 
in the AIPS task COMB (figure~\ref{spcindex}). 
Pixels with intensities less than the 3~$\sigma$ level in all the  
restored images are clipped. 
The spectral index images indicate 
that most parts of the two-side jet structure have optically 
thin synchrotron spectra at 15--43~GHz 
except at the inner edge;  
a steeply rising spectrum 
($\alpha^{22}_{15}=3.2\pm0.1$, $\alpha^{43}_{22}=3.1\pm0.1$; $S\propto\nu^{\alpha}$)
at the western edge of knot B and 
at the eastern edge of knot C3 are revealed. 
The spectral index exceeds the theoretical limit for synchrotron self-absorption ($\alpha=2.5$). 
The highly rising spectrum of the inner edge of the jets implies 
that the synchrotron emission is obscured 
through the free-free absorption (FFA) by the foreground dense plasma, 
and this is consistent with past multi-frequency observations 
(Kameno et al. 2001, Vermeulen et al. 2003 and Kadler et al. 2004).

\begin{figure}
\epsscale{.90}
\plotone{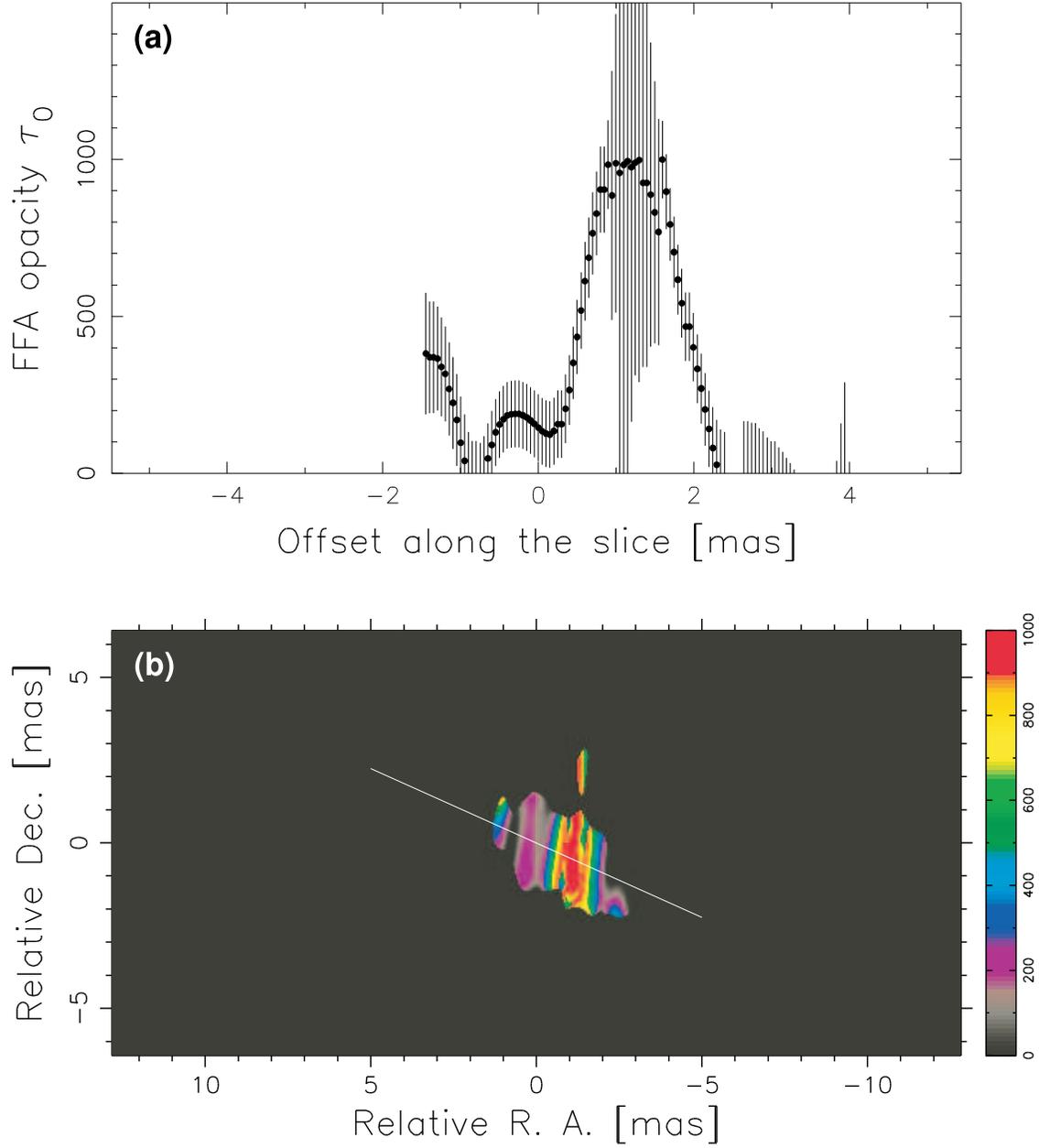}
\caption{
(a) Free-free absorption opacity along the jet axis (P.A. = 65$^{\circ}$). 
(b) Two dimension distribution of free-free absorption opacity. 
All images at 15, 22 and 43~GHz are restored with a same beam size of 1.30$\times$0.49 mas. 
The white line indicates the direction of slice cut.  
}
\label{opacity}
\end{figure}

We fitted the continuum spectrum at 15--43~GHz to FFA model, 
\begin{equation}
S_{\nu} = S_0 \nu^{\alpha} \exp{(- \tau_0 \nu^{-2.1})},
\end{equation}
where $S_0$ is the flux density in Jy at the frequency of 1~GHz 
extrapolated from the spectrum, 
and $\tau_0$ is the FFA opacity at 1~GHz
(Kameno et al. 2000). 
Obtained FFA opacity images along the jet axis  (P.A.=65$^{\circ}$) 
reveal that high opacity ($\tau_0 \sim 1000$)  is found in the inner edge
 (figure~\ref{opacity}). 
The fit in the inner edge has larger errors, 
because the continuum spectrum of the inner edge does not show 
the peak between 15 and 43 GHz. 
The space distribution of $\tau_0$ implies that 
the dense cold plasma covers $\sim$1~mas (0.1~pc), which is equal 
to the restoring beam size, 
in the inner edge 
of the jets, where the central engine is supposed to exist.

\subsection{H$_2$O maser emission}

\begin{figure}
\epsscale{.70}
\plotone{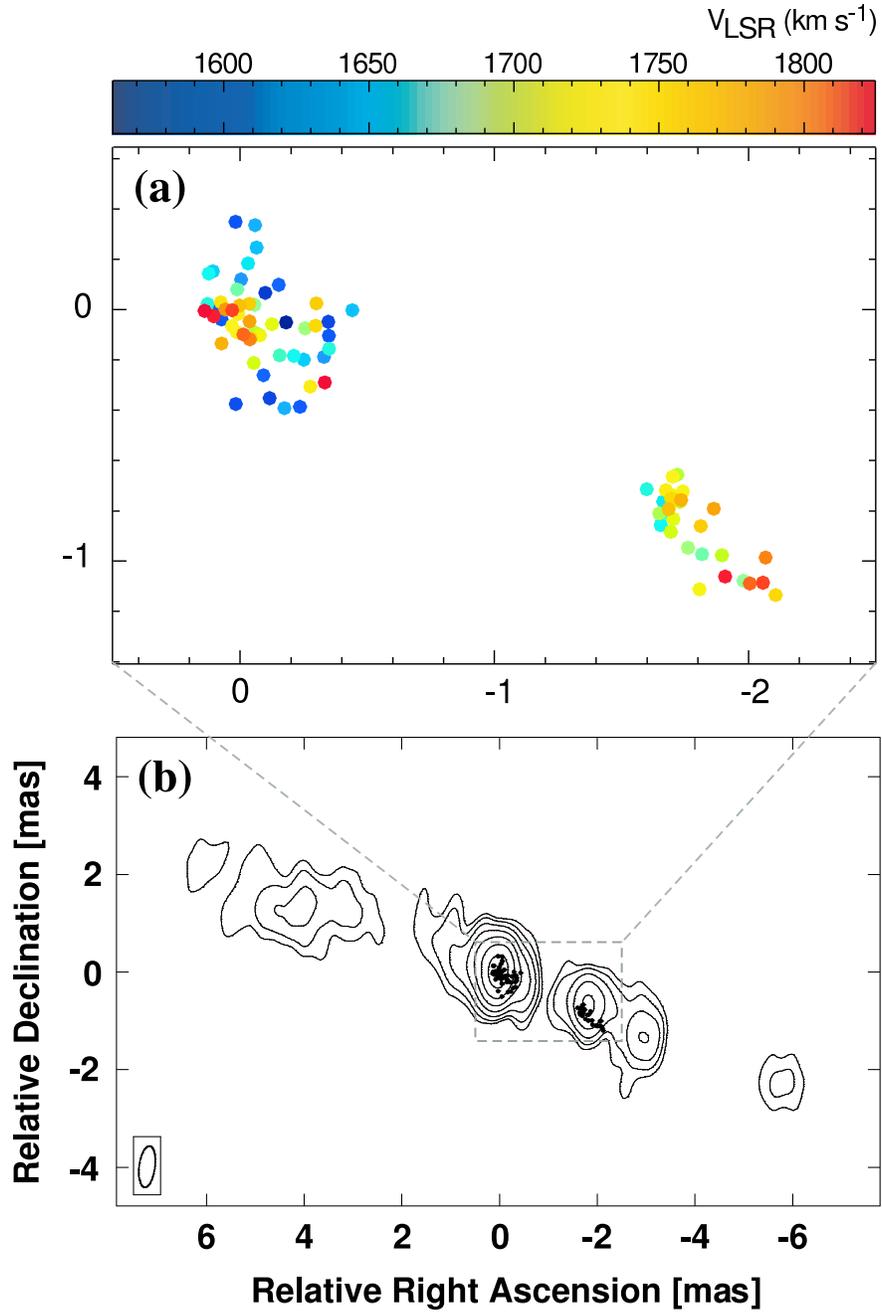}
\caption{(a) Doppler velocity distribution (colored filled circle) of 
H$_2$O maser spots in NGC~1052. 
(b) Relative distributions of H$_2$O maser spots (filled circle) 
with respect to the continuum image at 22~GHz (contour).
}
\label{maser}
\end{figure}

\begin{figure}
\epsscale{.90}
\plotone{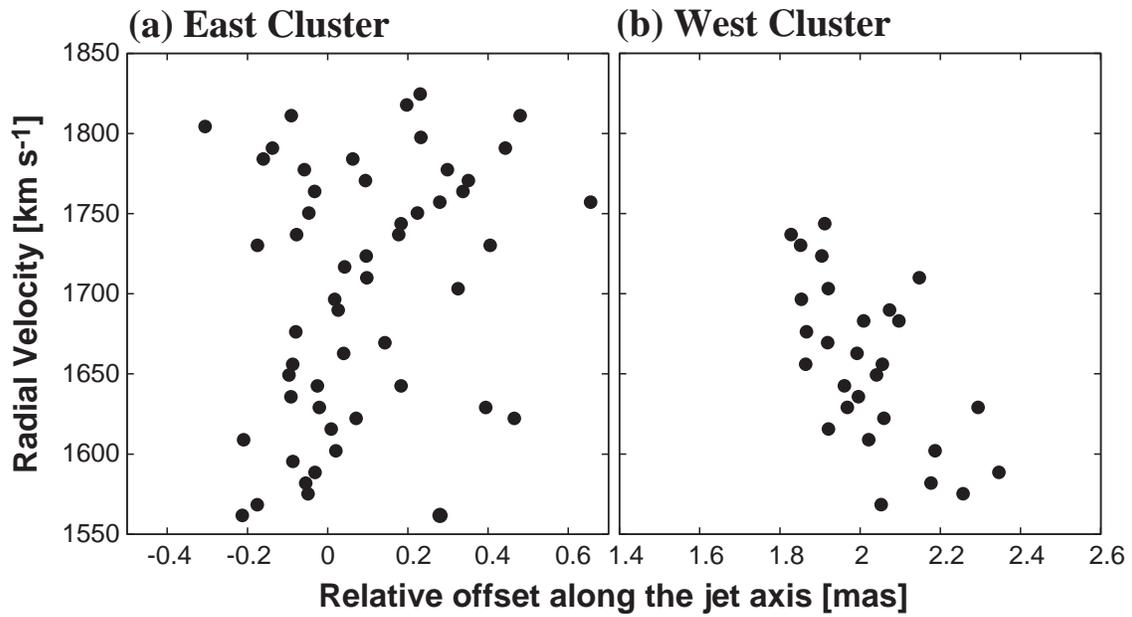}
\caption{Position-Velocity diagram of H$_2$O maser spots 
along the jet axis with respect to the peak of knot B.  
A clear velocity shift as a function of relative offset along the 
jet axis (P.A.=65$^{\circ}$ in figure~\ref{opacity}b) is seen in the western cluster (b). 
On the other hand, 
the eastern cluster seems to include several sub structures in velocity (a). 
}
\label{pv}
\end{figure}

In our observations, 
significant H$_2$O maser emission within the velocity range of 1550--1850 km~s$^{-1}$ 
were detected, 50--350 km~s$^{-1}$ redshifted from the systemic velocity of the galaxy. 
This is consistent with past single-dish observations 
(Braatz et al. 1994, Braatz et al. 1996, Braatz et al. 2003, Kameno et al. 2005). 
Figure~\ref{maser} shows the radio continuum image at 22~GHz and the distribution of H$_2$O maser spots of our observations (July 2000).     
The maser spots consist of two clusters; 
the eastern cluster and the western cluster are located 
on knots B and C3, respectively. 
The H$_2$O masers projected on the approaching jet, 
or the eastern cluster were detected at the first time. 
The velocity range is 
1550--1850 km~s$^{-1}$, which is same as the whole velocity width 
of the H$_2$O maser spectral profile.  
The maser spots in the eastern cluster 
with a velocity of 1550--1700 km~s$^{-1}$ 
are distributed within 1~mas (0.1~pc) of the knot B. 
The maser spots with a velocity of 1700--1850 km~s$^{-1}$ 
are more tightly concentrated within 0.2~mas (0.02~pc) 
of the peak of knot B. 
On the other hand, the western cluster is detected with a velocity range 
of 1550--1750 km~s$^{-1}$. 
These maser spots are distributed 
along the jet axis that span $\sim$ 1~mas (0.1~pc), 
and show some velocity shift along the direction. 
Position-velocity diagrams of the H$_2$O maser spots 
along the jet axis  (figure~\ref{pv}) also reveal the 
 trend of velocity gradient 
of $\sim$ 250 km~s$^{-1}$~mas$^{-1}$ 
in the western cluster. 
We note that Claussen et al. (1998) showed a velocity gradient 
along the east-west direction ($\sim$ 100 km~s$^{-1}$~mas$^{-1}$) 
in the maser cluster on the western jet knot. 
A velocity gradient in the eastern cluster is not obvious (figure~\ref{pv}a).  

\begin{figure}
\epsscale{.90}
\plotone{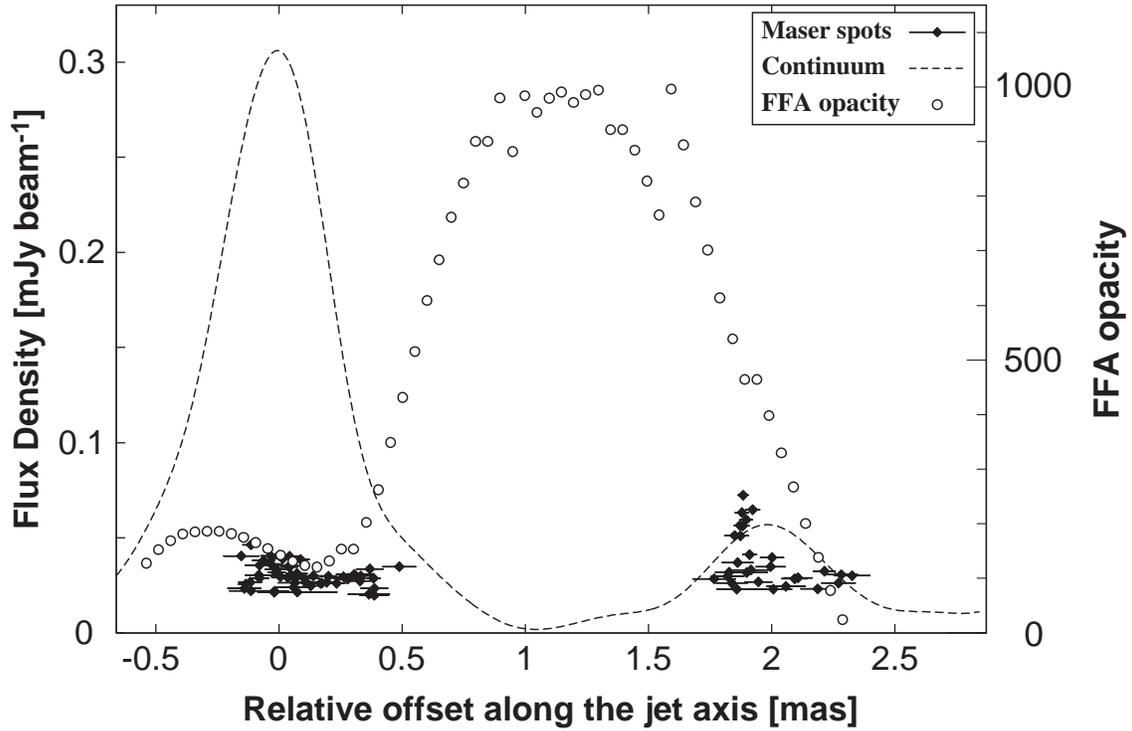}
\caption{The flux density of H$_2$O maser spots (filled diamond) 
and the continuum emission at 22 GHz (dashed line) 
along the right ascension offset, 
overlaid with the free-free absorption opacity along the right ascension. 
The continuum emission and the FFA opacity are obtained 
by the one-dimensional cut along the 
jet axis (P.A.=65$^{\circ}$ in figure~\ref{opacity}b), and is plotted along the right ascension offset. 
}
\label{slice}
\end{figure}

The peak flux density for each maser spot 
and the continuum emission at 22~GHz along the jet axis are shown in 
figure~\ref{slice}. 
The H$_2$O maser flux and the continuum flux do not appear to be correlated. 
The brighter maser spots in the western cluster are located
closer to the gap between knot B and C3.  
This trend is also seen in the plot of the flux density of H$_2$O maser spots vs. their right ascension offset in Claussen et al. (1988). 
In figure~\ref{slice}, the central engine may be located around 1 mas 
in relative right ascension, if the FFA opacity peak is a good indicator.

\section{Discussions}

\subsection{Orientation of the jet axis}

The brightness temperature ratio between the approaching and receding jets is  
related to the viewing angle of the jet axis and the true jet speed. 
Assuming that knots A and C1 form a symmetric pair of knots 
on either side of the nucleus, 
their intensity ratio $R$ is given by 
\begin{equation}
R=\frac{S_{A} \cdot \phi^{C1}_{maj} \cdot \phi^{C1}_{min}}{S_{C1} \cdot \phi^{A}_{maj} \cdot \phi^{A}_{min}}
= \biggl( \frac{1+\beta \cos{\theta}}{1-\beta \cos{\theta}} \biggr) ^{3-\alpha}
\end{equation}
where $S_A$ and $S_{C1}$ are the flux density of knots A and C1, respectively, 
$\phi^{A}_{maj}$, $\phi^{A}_{min}$  $\phi^{C1}_{maj}$ and $\phi^{C1}_{min}$ are the FWHM sizes of the knots 
A and C1, 
$\beta$ is the true jet velocity as a fraction of the speed of light ($v/c$), 
$\theta$ is the viewing angle of the jet axis, 
and $\alpha$ is the spectral index. 
Since the knots A and C1 are shifted 5 mas from the center,  
the effect of FFA on A and C1 would be small. 
Adopting $\beta=0.64$ (Kadler et al. 2004b) and $\alpha=-1$, 
 the viewing angle is estimated to be 79--80$^{\circ}$ and 76--90$^{\circ}$
 using the flux density at 15 and 22 GHz, respectively. 
 Therefore, the jet axis is considered to be nearly parallel to the sky plane. 
The estimated minimum angles are larger than that obtained 
by Kameno et al. (2001) and Vermeulen et al. (2003).  
In their data, the receding component located closer to the center, 
and  the emission of the receding component could be absorbed more 
due to the effect of FFA. 
The obtained intensities of the jets could be more asymmetry than the 
intrinsic intensities, and 
the asymmetric intensity ratio leads to the estimation of the inclined jet axis.

\subsection{Interpretation of H$_2$O maser}

There are two main explanations for what the H$_2$O masers of NGC~1052 are associated with: 
the jet or the circumnuclear torus (Claussen et al. 1998, Kameno et al. 2005). 
Here we discuss the plausible nuclear structure in NGC~1052,  
to account for the observed characteristics: 
(1) FFA plasma spanned $\sim$ 1~pc,  
(2) redshifted velocity of the H$_2$O masers with respect to the systemic velocity of NGC~1052, 
(3) H$_2$O masers appeared to be projected against knots B and C3, 
(4) velocity gradient of the H$_2$O masers along the jet,
and (5) more dominant FFA opacity and H$_2$O maser emission 
on the western receding jet as compared to the eastern approaching jet.

\begin{figure}
\epsscale{.90}
\plotone{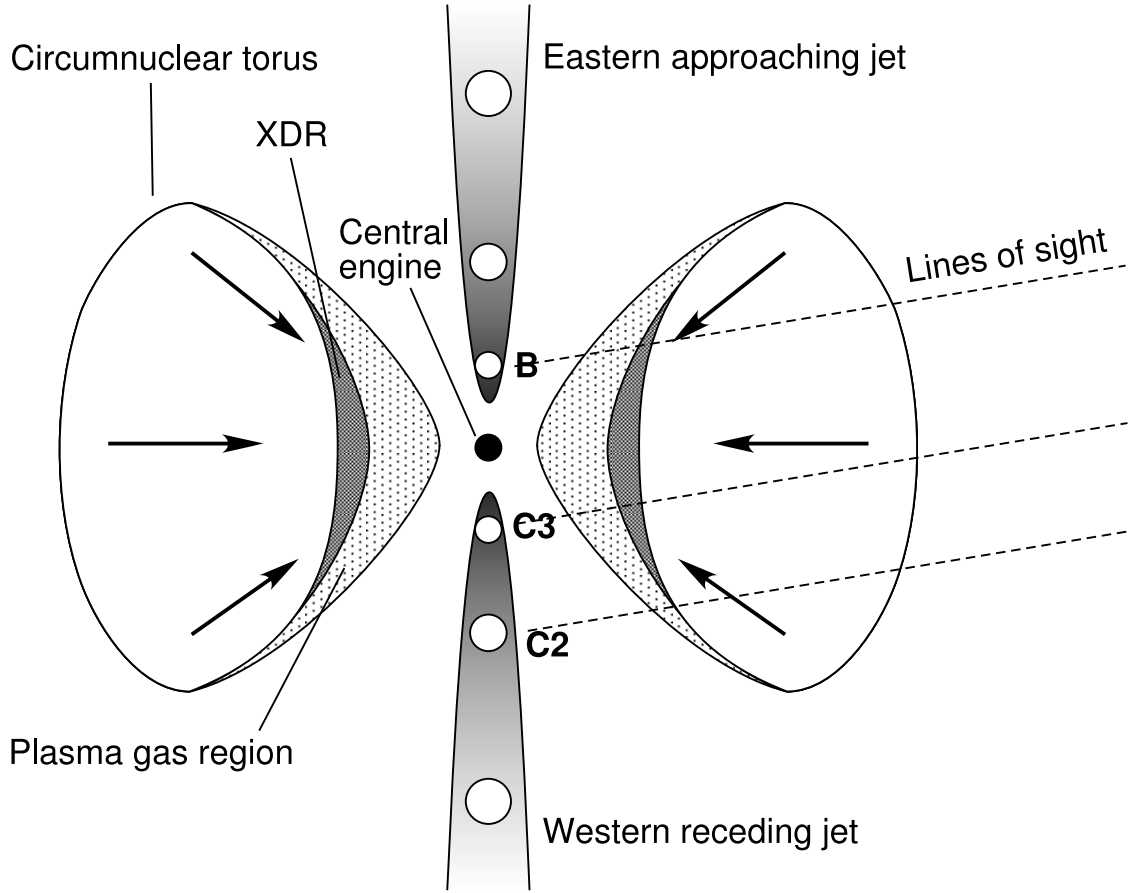}
\caption{
A cartoon showing the possible environment in the 
circumnuclear torus and jets in NGC~1052. 
Double-sided jet axis inclined by $\ge 76^{\circ}$ 
 with respect to the line of sight. 
Inner surface of the torus is ionized by X-ray emission.
X-ray Dissociation Region (XDR) is formed on the inner layer 
of the torus and amplify background continuum emission 
from the jet knot. 
On B and C3, we can see the H$_2$O maser emission and FFA absorption as well. 
On the other hand, only FFA appears on C2. 
Since the gas inside the torus is falling toward the central 
engine,  redshifted H$_2$O maser emission is detected.  
}
\label{torusmodel}
\end{figure}

One possible scenario is that the H$_2$O masers are associated with a 
circumnuclear torus, as illustrated in figure~\ref{torusmodel}. 
The H$_2$O masers are seen 
where the FFA opacity is large, and 
this suggests that the H$_2$O masers and the plasma exist close to 
each other.  
In the case of the H$_2$O megamaser emission in NGC~4258, 
Neufeld $\&$ Maloney (1995) and Herrnstein et al. (1996) 
proposed that the molecular disk consists of several layers including 
a heated molecular layer where the H$_2$O masers reside. 
Kameno et al. (2005) applied this idea to the circumnuclear 
torus model in NGC~1052. 
A hot ($\sim$ 8000 K) plasma layer is created on the inner surface of the torus because of the direct exposure to the X-ray radiation from the 
central source. 
This layer is responsible for the free-free absorption. 
The X-Ray Dissociation Region (XDR)  which lies immediately 
next to the plasma layer inside the torus, 
is heated above $\sim 400$~K as it is still partially irradiated 
by the X-ray radiation (Maloney 2002). 
Excited H$_2$O molecules 
in the XDR will amplify the continuum seed emission 
from the jet knots in the background and result in maser emissions. 
The presence of masers on both jets 
indicates the thickness of the torus along the orientation of the jet,   
covering at least knot B and C3. 
If the orientation of the jet axis is parallel to the sky plane, 
the thickness of the torus should be therefore 0.2~pc at least. 
More dominant FFA and H$_2$O masers on the receding jet 
support the circumnuclear torus scenario, 
since the path length within the torus toward the receding jet would be greater.

If the H$_2$O masers are associated with the torus, 
the redshifted spectrum of the H$_2$O maser emission
would be accounted for by a contraction toward the central engine. 
The positional-velocity diagram along the jet axis 
for each maser cluster (figure~\ref{pv}) 
gives the appearance 
that the H$_2$O maser gas closer to the central engine is more redshifted. 
Such a velocity shift as a function of positional offset 
could indicate the acceleration of the infalling gas 
toward the central engine. 
The blue-shifted H$_2$O maser emission would not be 
detected in the model.  
Because H$_2$O maser emission is seen 
when jet knots locate behind the XDR layer in the torus,  
the spatial structure of H$_2$O masers could be varied as the jet knots eject 
and run behind the torus. 
The maser gain length and the radial velocity along the 
line of the sight would change when the jet knots run behind the torus, 
and it could also cause time variations of maser in flux and velocity. 
This idea has been proposed by Kameno et al. (2005) 
to explain the emergence of a narrow maser feature seen in 2003 
at 1787 km~s$^{-1}$, 
and could also account for the variations in maser profile 
shown in Braatz et al. (2003). 

The outer region of the torus remains neutral either 
in molecular or atomic form.
Possibly, H{\scriptsize I} gas also could be associated 
with the outer region of the circumnuclear torus. 
It is reasonable to suppose that infalling H{\scriptsize I} gas 
toward the center results in the redshifted absorption line 
detected on both approaching and receding jets, 
just as in the case for the H$_2$O maser emission. 
Thus, the circumnuclear torus scenario can explain the observed characteristics.  

Another possible explanation is the excitation of the H$_2$O masers 
by the outflowing jet as Claussen et al. (1998) proposed,  
and as interpreted in the case of Mrk~348 (Peck et al. 2003). 
The jet excitation scenario is thought to have an advantage 
to explain the velocity gradient 
along the jet axis. 
The western maser cluster can be then explained easily 
because the maser spots in the western cluster are all redshifted 
and appear projected against the receding jet.   
However, the jet excitation scenario cannot explain 
why the masers of the eastern cluster 
projected against the approaching jet also shows the same redshifted spectrum. 
If the exact systemic velocity of the central engine is not 1490 km~s$^{-1}$ 
but around  1700 km~s$^{-1}$, the jet excitation scenario would become likely.  
For the masers in the western cluster moving with the most redshifted velocity 
($\sim$ 400 km~s$^{-1}$) from the systemic velocity of the galaxy, 
the maser gas should move 1.9$\times$10$^{-3}$ pc eastward 
from November 1995 to July 2000, 
because the jet axis is close to the sky plane.  
This motion is too small to detect for the five-year multi-epoch observations.  
Further VLBI observations are necessary 
in order to detect the proper motion of H$_2$O maser spots 
in the jet excitation scenario. 

\section{Conclusions}

We conducted multi-frequency observations toward the center of NGC 1052 
at 15, 22 and 43 GHz with the VLBA.   
Ratios of the brightness temperature of the approaching and the receding jet knots indicate that 
the angle of the jet axis to the line of sight is $>76^{\circ}$, 
or nearly parallel to the sky plane. 
The aligned continuum images show 
steeply rising spectra spanned in the inner 0.2 pc around the nucleus, 
which imply FFA by dense plasma in a circumnuclear torus. 
H$_2$O maser emission is detected at  velocities 
by 150--350 km~s$^{-1}$ redshifted
from the systemic velocity of the galaxy, 
which is consistent with past single-dish observations.  
The maser spots are projected against the inner knots of 
both the approaching and receding jets, where the FFA opacity is large. 
A clear velocity gradient along the jet axis in the western cluster is seen, 
but the eastern cluster does not show such a clear velocity trend.  
The more redshifted maser spots lie closer to the center. 

Positional coincidence between the H$_2$O masers  
and a plasma torus suggests 
that the H$_2$O maser emission arises from the circumnuclear torus. 
If the H$_2$O masers is moving as contraction toward the center, 
the redshifted spectrum and the velocity gradient of the H$_2$O maser emission can be explained fairly well.  
Alternatively, the H$_2$O masers could be excited 
by the interaction between the jet and circumnuclear molecular clouds.  
A weak point of the jet excitation scenario is the difficulty to explain 
why the eastern cluster of H$_2$O masers projected 
against the approaching jet is also redshifted.







\acknowledgments
We are grateful to Kenta Fujisawa, Kiyoaki Wajima for their helpful comments.
We also appreciate Paul T. P. Ho for his critical reading of the manuscript and 
invaluable discussions.
The VLBA is operated by the NRAO, a facility of the National Science Foundation operated under cooperative agreement by Associated Universities, Inc. 
This research was partially supported by the Ministry of Education, Science, Sports and Culture, Grant-in-Aid for Young Scientists (B), No. 17740115.

\clearpage

\begin{table}
\begin{center}
\caption{Beam size and rms noise level of the images}
\begin{tabular}{lcccc} \\ \tableline \tableline 
    $\nu$ & Major axis & Minor axis & P.A. & rms level  \\
    (GHz) & (mas) & (mas) & (deg.) & (mJy~beam$^{-1}$) \\ \tableline
    15           &  1.30  &  0.49 & -4.8 & 0.24 \\
    22$^{a}$ &  0.86  &  0.32 & -7.1 & 1.07 \\
    22$^{b}$ &  0.86  &  0.32 & -7.1 & 5.66 \\
    43   &  0.39  &  0.17 & -2.8 & 1.45 \\
\tableline 
\end{tabular}
\tablenotetext{a}{Continuum map.  }
\tablenotetext{b}{Channel map of H$_2$O maser. Velocity resolution is 6.74 km~s$^{-1}$.}
\end{center}
\label{beam}
\end{table}

\begin{table}
\begin{center}
\caption{Parameters of Jet Knots by Gaussian Fitting}
\begin{tabular}{lcccccc}
\tableline\tableline
$\nu$ & Knot & Flux Density & Relative R.A.\tablenotemark{a} & Relative DEC\tablenotemark{a} & Major\tablenotemark{b} & Minor\tablenotemark{b} \\
 (GHz)   &   & (mJy)  & (mas)  & (mas)  & (mas)  & (mas) \\
\tableline
15 {......}	& A	& 330$\pm$5	& 4.066$\pm$0.006	& 1.328$\pm$0.004	& 3.0		& 1.6 \\
	& B	& 639$\pm$1	& $\cdots$	& $\cdots$		& 1.3		& 0.6 \\
	& C3	& 83$\pm$1	& -1.985$\pm$0.001	& -0.774$\pm$0.002	&	1.3 & 0.8 \\
	& C2	& 121$\pm$1	& -3.049$\pm$0.001	& -1.297$\pm$0.002 &	1.4 & 0.8 \\
	& C1	&  95$\pm$2	& -6.014$\pm$0.007	& -2.382$\pm$0.006	& 1.8		& 1.5 \\ 
	& D	& 17$\pm$2	& -13.53$\pm$0.06	& -6.17$\pm$0.05	& 2.1		& 1.7 \\
22 {......}	& A	& 129$\pm$10	& 4.09$\pm$0.07	& 1.35$\pm$0.04 	& 2.1		& 1.2 \\
	& B	& 560$\pm$3	& $\cdots$		& $\cdots$		& 0.9		& 0.6 \\
	& C3	& 119$\pm$3	& -1.866$\pm$0.007	& -0.698$\pm$0.007	& 0.9		& 0.7 \\
	& C2	& 68$\pm$4	& -2.97$\pm$0.01	& -1.32$\pm$0.02	& 1.1		& 0.7 \\
	& C1	& 26$\pm$4	& -5.92$\pm$0.07	& -2.30$\pm$0.08	&  1.0	& 0.9 \\
43 {......}	& B	& 305$\pm$4	& $\cdots$		& $\cdots$		&  0.4	& 0.3 \\
	& C4	& 106$\pm$3	& -0.605$\pm$0.003	& -0.245$\pm$0.004	&  0.4	& 0.3 \\
	& C3	& 201$\pm$7	& -1.45$\pm$0.01	& -0.611$\pm$0.007 &  0.8	& 0.4 \\
	& C2	& 25$\pm$4	& -3.05$\pm$0.01	& -1.33$\pm$0.04	&  0.7	& 0.2 \\
	& C1	& 10$\pm$2	& -6.35$\pm$0.01	& -2.55$\pm$0.02	&  0.4	& 0.2 \\
\tableline
\end{tabular}
\tablenotetext{a}{Relative position with respect to knot B}
\tablenotetext{b}{FWHM of the Gaussian component}
\end{center}
\label{gaussian}
\end{table}

\end{document}